\documentclass[final,1p,times]{elsarticle} 
\usepackage{graphicx} 
\usepackage{amssymb} 
\usepackage{amsthm} 
\usepackage{lineno} 

\journal{Nuclear Physics A} 

\usepackage{epsfig,epsf}
\usepackage{verbatim}
\usepackage{epstopdf}
\usepackage{bm} 
\newcommand{\beq}{\begin{equation}}
\newcommand{\beql}[1]{\begin{equation}\label{#1}}
\newcommand{\eeq}{\end{equation}}
\newcommand{\bea}{\begin{eqnarray}}
\newcommand{\eea}{\end{eqnarray}}
\newcommand{\be}{\begin{eqnarray}}
\newcommand{\ee}{\end{eqnarray}}
\def\eq#1{{(\ref{#1})}}
\def\fig#1{{Fig.~\ref{#1}}}

\newcommand{\as}{\alpha_s}

\def\b#1{\mathbf{#1}}

\begin{document} 

\begin{frontmatter} 


\title{Breakdown of $k_T$-factorization and $J/\psi$ production in dA collisions }

\author{Kirill Tuchin}

\address{Department of Physics and Astronomy, Iowa State University, Ames, IA 50011\\ and \\
RIKEN BNL Research Center, Upton, NY 11973-5000}

\begin{abstract} 

In spite of the sweeping coherence effects in high energy hadron and nuclei collisions,  $k_T$-factorization can be recovered for the inclusive gluon production in $pA$ collisions at the leading logarithmic order. In open charm production at RHIC $k_T$-factorization holds numerically with about 10--20\% accuracy. This allows to extrapolate the cold nuclear matter effect observed in light and charm meson production in $dA$ collisions to that in $AA$ ones. Unlike the open charm, the breakdown of factorization in $J/\psi$ production is severe. Indeed, already at the lowest order in gluon density the main contribution to the inclusive cross section is proportional to the square of gluon density in the nucleus. As a consequence, one cannot infer the cold nuclear matter effect on $J/\psi$ production in $AA$ collisions from that in $dA$. We present the calculation of $J/\psi$ multiplicity in the framework of the CGC (color glass condensate)/saturation and show that it agrees with the experimental data.

\end{abstract} 

\end{frontmatter} 

\linenumbers 



\section{Introduction}
One of the cornerstones of the hard perturbative QCD (hpQCD) is factorization theorems, which state that soft non-perturbative part of the scattering cross section for any hard process can be encoded in universal parton distribution functions (pdf's). `Hard' means that the momentum transfer $Q$ is much larger than any intrinsic hadron momentum scale. 
According to Gribov, Levin and Ryskin \cite{Gribov:1984tu}, at high energies there is a universal scale characterizing hadronic wave functions -- the saturation momentum $Q_s$ -- that rapidly increases  as a power of energy.  One therefore expects that at higher energies  factorization is broken down in a wide region of momenta $p_T\lesssim Q_s$. Understanding the structure of  inclusive processes in the region where the hpQCD factorization is not applicable is important for quantifying the role of the cold nuclear matter effects in $pA$ and $AA$ collisions. In this article I review our  
present understanding of the subject and its implications for inclusive $J/\psi$ production.

\section{Factorization in inclusive gluon and quark production}

Factorization in inclusive gluon production in $pA$ collisions  in the low $x$ region was investigated in \cite{Kovchegov:2001sc}, where the cross section was derived that  re-sums all leading logarithmic   contributions $\as\ln(1/x)\sim 1$ (LLA)  for a heavy nucleus  in the quasi-classical limit $\as^2A^{1/3}\sim 1$.  One does not expect any hpQCD factorization to apply in this case because higher twist interactions  of valence quarks and gluons give contributions of order unity. Nevertheless, despite the fact that individual diagrams break factorization in covariant and light-cone gauges, the final re-summed expression can be cast in the $k_T$-factorized form. Unlike in hpQCD, the physical quantity that is factorized -- the unintegrated gluon distribution $\varphi(x,Q^2)$ -- is not soft and can be calculated perturbatively owing to existence of a hard scale $Q_s\gg \Lambda_\mathrm{QCD}$.  Another surprising fact is that contrary to naive expectations $\varphi(x,Q^2)$ is related not to the momentum space Fourier-image of the nucleus gluon field correlation function $\langle\b A(\b 0)\cdot \b A(\b x)\rangle$, but rather to the Fourier-image of $\nabla^2_r N(\b r, \b b,y)$, where $N(\b r, \b b,y)$ is the imaginary part of the forward elastic scattering amplitude of a color dipole of size $\b r$ at impact parameter $\b b$ and rapidity $y=\ln (1/x)$ in the heavy nucleus. We will see that although the inclusive gluon production in $pA$ collisions is the only known case were  $k_T$-factorization survives, factorization of the multipoles in the transverse coordinate space is the general feature of the low-$x$ cross sections. It must be stressed that this multipole factorization does not imply hpQCD factorizations ($k_T$ or collinear ones) and neither opposite is generally true. A $k_T$-factorization formula derived  in \cite{Kovchegov:2001sc} led to successful phenomenology of inclusive hadron production in $dA$ collisions at RHIC were the suppression of hadrons at forward rapidities and the Cronin enhancement at mid-rapidity were qualitatively predicted \cite{Kharzeev:2003wz,Albacete:2003iq} and then quantitatively described in the CGC framework \cite{Kharzeev:2004yx}. \footnote{An alternative approach that incorporates the coherence effects, but attributes the observed suppression to the peculiarities of the fragmentation process is presented in \cite{Kopeliovich:2005ym}. The two mechanisms can be discriminated by observing whether the suppression pattern scales with $x_{Au}$ (CGC) or $x_{D}$ (fragmentation).}

 $k_T$-factorization in $pA$ collisions does not imply any factorization in $AA$. 
In fact, no rigorous analytic result for an inclusive process exists in the latter case. As in $pA$, individual diagrams in any gauge break the hpQCD factorization. It is however possible that the re-summed result exhibit a simpler structure than the individual diagrams. Thus, using a theoretically motivated conjecture it has been shown in \cite{Kovchegov:2000hz} that inclusive gluon production in $AA$ can be written in a form that breaks  $k_T$-factorization only logarithmically. Based on this conjectured approximate $k_T$-factorization, one can derive an important qualitative conclusion about the observed strong suppression of inclusive hadron production in $AA$ collisions at RHIC. Since the analogous process in $pA$ collisions in  the same kinematic region exhibits Cronin enhancement, we conclude that the suppression in $AA$ is not a cold nuclear matter effect. We see that factorization plays a key role in establishing existence of a new form of nuclear matter produced in $AA$ collisions. 

Production of heavy quark in $pA$ collisions at low-$x$ was calculated in 
\cite{Gelis:2003vh,Tuchin:2004rb,Blaizot:2004wv,Kovchegov:2006qn}. One expects that  the hpQCD factorization is  applicable if the saturation momentum is much smaller than the quark mass $m$ \cite{Kharzeev:2003sk}. At RHIC $Q_s\sim m$ for charm and bottom, hence factorization is broken is both cases. Indeed, analysis of \cite{Fujii:2005vj} indicates that the semi-classical calculations of \cite{Gelis:2003vh,Blaizot:2004wv} disagree with  $k_T$-factorization by 10-20\% at 
the $t$-channel gluon transverse momenta around $m$. A common feature of the inclusive gluon and heavy quark production is that at transverse momenta much higher than the saturation momentum, cross sections reduce to the  LO hard perturbative ones and consequently  factorize. In other words,  the hpQCD factorization is restored in the kinematic region where the operator product expansion  is applicable. It is important therefore that the leading term in the twist expansion coincides with that of hpQCD. This is not so in the case of the $J/\psi$ production.

\section{Factorization breakdown for inclusive $J/\psi$ production in nuclear collisions}

The mechanism of $J/\psi$ production in high energy nuclear collisions is different from that in hadron-hadron collisions \cite{KT,Kharzeev:2008cv,Kharzeev:2008nw}. Consider first the $J/\psi$ production in $pp$ collisions. At high energies  formation time of the $J/\psi$ wave function is much larger than the size of the interaction region. Indeed, in the nucleus rest frame, the former is $2M_\psi l_c/(M_{\psi'}-M_\psi)$ where $l_c= 1/(xM_N)>R_A$ is the coherence length and $M_{\psi'}-M_\psi\ll M_\psi$, while the latter is $R_A$. Therefore, we need to take into account only interaction of the $c\bar c$ pair with the nucleus. This interaction however depends on the quantum state of the $c\bar c$ pair. In the color singlet model it is in the $1^{--}$ color singlet state while in the color evaporation model it can be in any state with invariant mass below the $D$-meson threshold.  Although the color singlet  model is physically well-motivated, it underestimates the $J/\psi$ yield.  We therefore assume that $c\bar c$ is produced in the $1^{--}$ state with any color. Since none of the existing approaches to $J/\psi$ production (including the non-relativistic QCD model) in $pp$ collisions agrees with all the available data, one may doubt the applicability of our model for $J/\psi$ production in $pA$ collisions. Note, however, that there are indications that  solution to the $J/\psi$ production puzzle lies in understanding the  role of the higher twist contributions \cite{jwq}. In heavy nuclei, where $\as^2A^{1/3}\sim 1$, only certain higher twists are enhanced -- namely those  corresponding to interaction of the color dipoles with different nucleons. Therefore, higher twist contributions in $pA$ collisions can be taken into account in a systematic way. 

\begin{figure}[ht]
      \includegraphics[width=10cm]{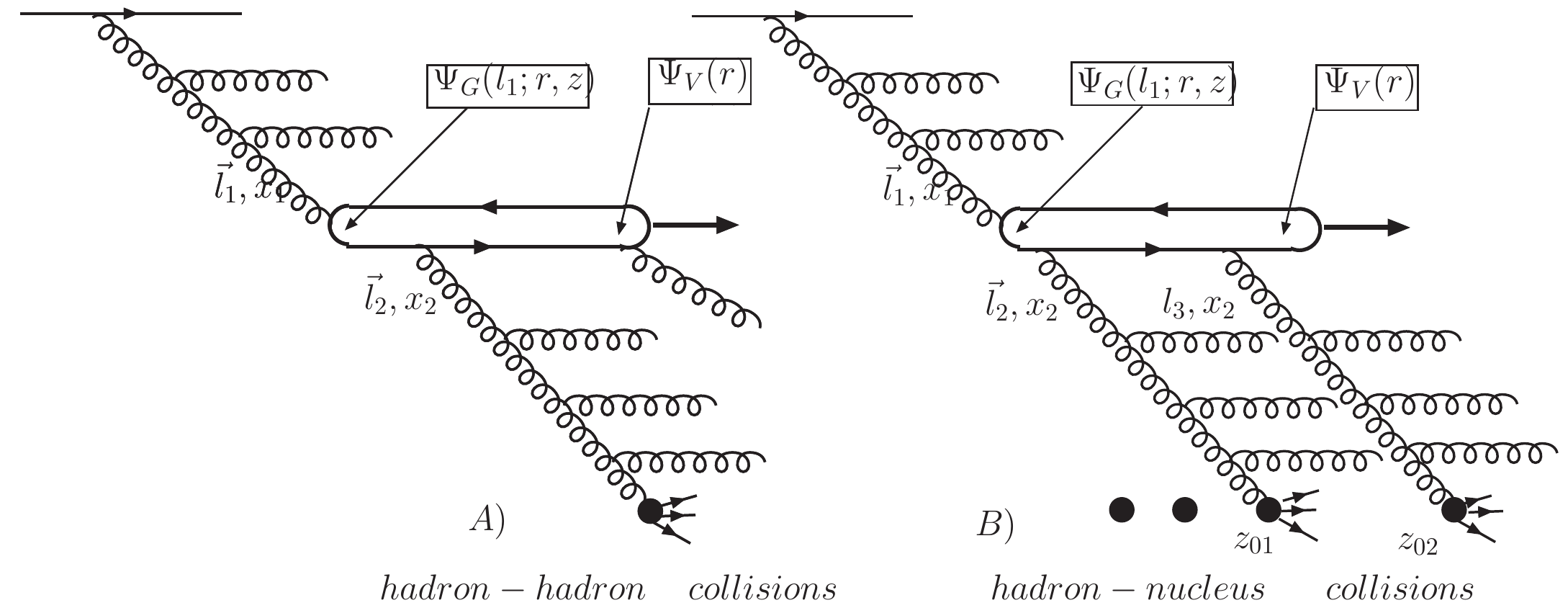} 
\caption{The  process of inclusive $J/\psi$  production in (A) hadron-hadron  and (B)  hadron-nucleus collisions. }
\label{psi1}
\end{figure}

Consider now two possible production mechanisms at the lowest order in $\as$ illustrated in  \fig{psi1}: (A) $g+g\to J/\psi +g$, twhich is of the order $\mathcal{O}(\as^5 A^{1/3})$ and  (B) $g+g+g\to J/\psi$, which is of the order  $\mathcal{O}(\as^6 A^{2/3})$. Since $\as^2A^{1/3}\sim 1$, the mechanism in \fig{psi1}-B is parametrically enhanced. 
Notice, that this leading contribution explicitly breaks $k_T$-factorization as it is proportional to $xG(x_1)[xG(x_2)]^2$.

At high energies, when $l_c\gg R_A$, all possible scatterings of the $c\bar c$ pair inside the nuclear medium must be taken into account. The number of scatterings on each diagram is restricted  however  by the requirement that the $c\bar c$ pair is in $1^{--}$ state. In other words, the number of inelastic interactions of the charm dipole in the nucleus 
must be even. The corresponding scattering amplitude reads
\beql{inel}
T_\mathrm{in}(\b r, \b r')= e^{-r^2 Q_s^2/8}\,e^{-r'^2 Q_s^2/8}
\left( \cosh[2\b r\cdot \b r'\, Q_s^2/8]-1\right)\,,
\eeq
where $r$ and $r'$ are the $c\bar c$ dipoles in the amplitude and in the complex conjugated one. Of course, there is no such restriction on the elastic amplitude involving exchange of gluon pair with the vacuum quantum numbers. The elastic amplitude is  given by 
\beql{elas}
T_\mathrm{el}(\b r, \b r')= \left( 1-e^{-r^2 Q_s^2/8}\right)\left( 1-e^{-r'^2 Q_s^2/8}\right)\,.
\eeq
To obtain the cross section one has two convolute these amplitudes with the virtual gluon and $J\psi$ wave functions, given in  e.g.\ \cite{Kovchegov:1999kx,Gotsman:2003ww}.

Experimental data is expressed in terms of the nuclear modification factor  defined as
\beql{nmf}
R_{AB}= \frac{\int_\mathcal{S}\, d^2b\frac{d\sigma^J_{AB\to JX}}{dy\, d^2b}}{A\,B\,\frac{d\sigma_{pp\to JX}}{dy}}\,,
\eeq
 where $\mathcal{S}$ stands for the overlap area of two colliding nuclei. We calculate the $J/\psi$ production in $pp$ collisions according to \fig{psi1}A by replacing $T_\mathrm{in}$ in \eq{inel} with 
 \beql{inelcc}
T_\mathrm{in}'(\b r, \b r')= e^{-(\b r -\b r')^2 Q_s^2/8}- e^{-r'^2 Q_s^2/8}e^{-r^2 Q_s^2/8}\,,
\eeq
 where no restriction on the number of inelastic interactions has been made. Note also that the overall normalization is still not determined since the $pp$ cross section is proportional to the probability of soft gluon emission. We approximate this probability by a constant. 
 
\begin{figure}[ht]
\begin{center}
      \includegraphics[width=6.5cm]{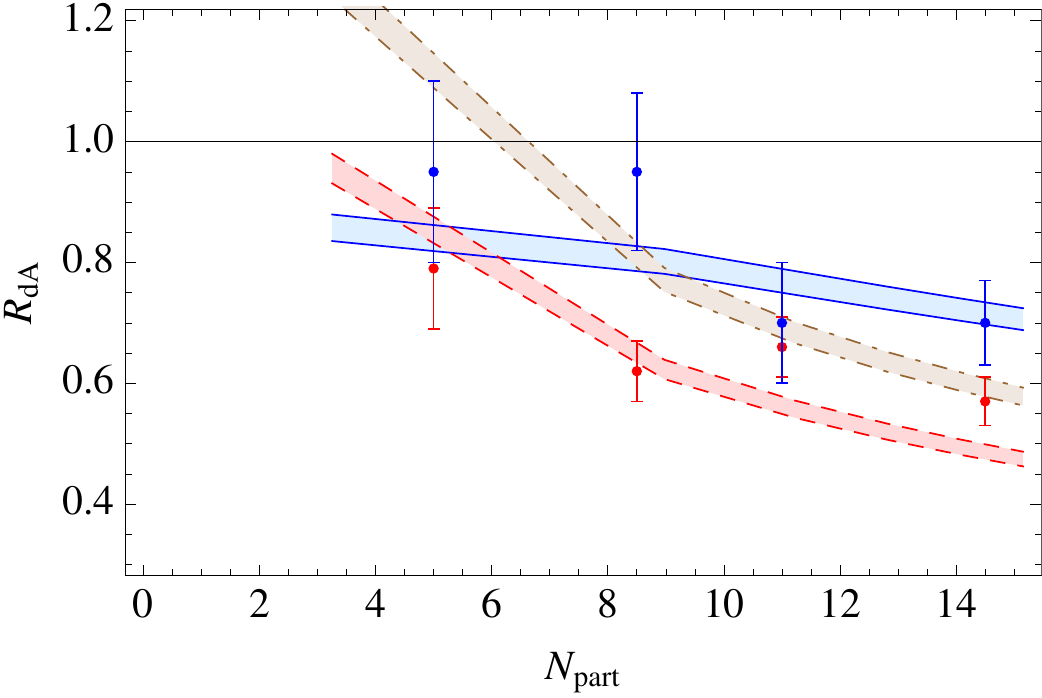} 
 \end{center}     
\caption{Nuclear modification factor for $J/\psi$ production in heavy-ion collisions for different rapidities. Solid (blue) line corresponds to  RHIC $y=0$, dashed (red) lines -- RHIC $y=1.7$, Dot-dashed (brown) line -- LHC mid-rapidity. Experimental data from \cite{Adler:2005ph}.}
\label{fig:BdNdy}
\end{figure}

 To take into account the low-$x$ quantum evolution we recall that the initial condition for the BK \cite{Balitsky:1995ub,Kovchegov:1999yj} evolution equation is given by the Glauber--Mueller formula for the forward dipole--nucleus elastic scattering amplitude \cite{Mueller:1989st}
\beql{init.cond}
N(\b r, \b b, y_0)= 1-e^{-\frac{1}{8}\b r^2Q_{s}^2(y_0)}\,.
\eeq
Therefore, we can incorporate the evolution effects by writing the scattering amplitudes \eq{inel},\eq{elas},\eq{inelcc} in terms of the the amplitude $N$ and letting the latter depend on rapidity $y$ as dictated by the BK equation. Energy evolution of $N(\b r, \b b, y)$ is taken into account using the  KKT model \cite{Kharzeev:2004yx} which parameters are fixed to describe the inclusive hadron production in $pA$ collisions. The result of the calculation is shown in \fig{fig:BdNdy}.
We observe that a good agreement with experimental data. We expect that the breakdown of  the hpQCD factorization has important impact on the $J/\psi$ production in $AA$ collisions. Because of this the magnitude of the nuclear modification factor $R_{AA}$ cannot be inferred from the $dA$ calculations.

\section{Conclusions}\label{concl}

We discussed the breakdown of the hard perturbative factorization in the gluon saturation region. This happens due to coherence effects over the entire nuclear length. This effect invalidates approaches that use  pdf's parametrized to include the ``nuclear shadowing" corrections. The very notion of pdf as a universal quantity is not applicable for processes with typical momentum transfer of the order of the saturation momentum. Fortunately,  rigorous analytical perturbative results can be derived for processes involving scattering off heavy nuclei. In particular, a new mechanism of $J/\psi$ production in $pA$ collisions has been suggested. It predicts dependence on  nuclear weight $A$ and rapidity $y$ which  is in a good agreement with the experimental data. We plan to extend our calculation to heavy-ion collisions and also calculate the $J/\psi$ polarization. Our approach can be applied to $pp$ collisions as a model resummation of higher twists. 


\section*{Acknowledgments} 
I am grateful to D.~Kharzeev, Yu.~Kovchegov, E.~Levin and  M.~Nardi for a fruitful collaboration and to J.W.~Qiu for informative discussions. 
This work was supported in part by the U.S. Department of Energy under Grant No. DE-FG02-87ER40371; I would like to
thank RIKEN, BNL,
and the U.S. Department of Energy (Contract No. DE-AC02-98CH10886) for providing facilities essential
for the completion of this work.  RBRC-797.

\end{document}